\documentclass[5p,twocolumn]{ifacconf}

\usepackage{indentfirst}
\usepackage{array}
\usepackage{natbib}
\usepackage{graphicx}
\usepackage{epstopdf}
\usepackage{color}
\usepackage[cmex10]{amsmath}
\usepackage{amsfonts,amssymb,amsxtra,balance}
\usepackage{bm}
\usepackage{mathrsfs}
\usepackage{calc,ifthen}
\usepackage{mathrsfs}
\usepackage{amssymb}

\usepackage{subfig}
\usepackage{caption}

\usepackage{tikz}
\usepackage{circuitikz}

\newtheorem{assumption}{Assumption}

\newtheorem{proposition}{Proposition}

\newtheorem{lemma}{Lemma}
\newtheorem{remark}{Remark}

\def\hal{{1 \over 2}}

\def\Linf{{\cal L}_\infty}
\def\Lt{{\cal L}_2}
\def\L2e{{\cal L}_{2e}}

\def\hal{{1 \over 2}}

\def\begequarrs{\begin{eqnarray*}}
\def\endequarrs{\end{eqnarray*}}
\def\begequarr{\begin{eqnarray}}
\def\endequarr{\end{eqnarray}}
\def\begarr{\begin{array}}
\def\endarr{\end{array}}
\def\begequ{\begin{equation}}
\def\endequ{\end{equation}}
\def\lab{\label}
\def\begdes{\begin{description}}
\def\enddes{\end{description}}
\def\begenu{\begin{enumerate}}
\def\begite{\begin{itemize}}
\def\endite{\end{itemize}}
\def\endenu{\end{enumerate}}

\def\lef[{\left[\begin{array}}
\def\rig]{\end{array}\right]}

\def\begcen{\begin{center}}
\def\endcen{\end{center}}
\def\begrem{\begin{remark}\rm}
\def\endrem{\end{remark}}
\def\begcas{\begin{cases}}
\def\endcas{\end{cases}}

\begin{document}
\begin{frontmatter}
	\title{ Global Tracking Passivity--based PI Control of Bilinear Systems and its Application to the Boost and Modular Multilevel Converters}
	\author[first]{R. Cisneros}
	\author[second]{M. Pirro}
	\author[third]{G. Bergna}
	\author[first]{R. Ortega}
	\author[second]{G. Ippoliti}
	\author[third]{M. Molinas}
	\address[first]{Laboratoire de Signaux et Syst\'emes, Sup\'elec, 91190 Gif-sur-Yvette, France (e-mail: cisneros[ortega]@lss.supelec.fr)}
	\address[second]{Dipartimento Ingegneria dell'Informazione, Universit\`a Politecnica delle Marche, 60121 Ancona, Italia (e-mail:m.pirro[gianluca.ippoliti]@univpm.it)}
	\address[third]{Norwegian University of Science and Technology, 7491 Trondheim, Norway (e-mail: gilbert.bergnadiaz@supelec.fr, marta.molinas@ntnu.no)}
 \begin{abstract}
This paper deals with the problem of  trajectory tracking of a class of bilinear systems with time--varying measurable disturbance, namely, systems of the form $\dot{x}(t)=[A+\sum_{i}u_{i}(t)B_{i}]x(t)+d(t)$.
A set of matrices $\{A,B_i\}$ has been identified, via a linear matrix inequality, for which it is possible to ensure global tracking of (admissible, differentiable) trajectories with a simple linear time--varying PI
controller. Instrumental to establish the result is the construction of an output signal with respect to which the incremental model is passive. The result is applied to the boost  and  the  modular multilevel
converter for which experimental results are given.
\end{abstract}

\end{frontmatter}
\section{Introduction}
Bilinear systems are a class of nonlinear systems that describe a broad variety of physical and biological phenomena \cite{intro1} serving, sometimes, as a natural simplification of more complex nonlinear
systems. There is an amount of literature devoted to the study of the intrinsic properties or to stabilization of equilibrium points for these systems, see for example \cite{intr2}. However, to the best of our
knowledge, there is no general result for the design of controllers that ensure {\em global tracking} of (admissible, differentiable) trajectories.

The main objective of this paper is to provide a theoretical framework---based on the property of passivity of the incremental model---to establish such a result. Our motivation to pursue a passivity framework
is twofold, on one hand, it encompasses a large class of physical systems. On the other hand, it naturally leads to the design of PI controllers, which are known to be simple, robust and widely accepted by
practitioners. Our main result is an extension, to the problem of tracking trajectories, of \cite{JAYetal,SANVER} that treat the regulation case. See also \cite{CASetal} for its application to PI stabilization
of RLC circuits and \cite{intr3} where the result is used in power converters.

An important motivation for our research is the derivation of simple tracking controllers for power converters. In classical applications of power converters the control objective is to {\em regulate} the
output voltage (or current) around some constant desired value. Modern applications, on the other hand, are concerned with the more demanding specification of ensuring an effective transfer of power between
the sources and the loads---an objective that translates into the task of {\em tracking time--varying references}.

In the paper our theoretical result has been illustrated with the application to two important problems arising in power electronic systems. The first one is the problem of Power Factor Compensation (PFC), which
arises in renewable energy and motor drive systems with stringent specifications on efficiency, harmonic distortion and voltage regulation \cite{ElFadili2012230,6162764,5482119,cimini2013passivity}. The second
problem is related to High Voltage Direct Current (HVDC) transmission, which has recently attracted a lot of interest \cite{HVDCoverview,abcHVDC,Pros,6862419}. HVDC transmission consists of a grid
comprising mostly DC lines that integrates, via voltage source converters, renewable energies from distant locations. One of the most promising candidates to integrate the topology of the grid is the Modular
Multilevel Converter (MMC) \cite{MMC}, which has several advantages with respect to its predecessors, such as high modularity, scalability and lower losses. In view of its complicated topology and operating
regimes, controlling the MMC is no simple task. In particular, to exploit the full potential of the MMC it seems to be necessary to develop control strategies for the system operating in the rotating ($abc$)
frame \cite{gbJour2}, this is contrast with the classical strategies developed in fixed ($dq0$) frames \cite{CCSC,gbJour}.  This situation leads to a tracking problem instead of the typical regulation one. 
The tracking problem of bilinear systems has been addressed within the context of switched power converters.  In \cite{intr5} a methodology to track periodic signals for non-minimum phase boost converters
based on a stable inversion of the internal dynamics taking the normal form of an Abel ordinary differential equation was presented---see also \cite{intr6}. There are also schemes involving sliding mode
control, for example \cite{intr10,intr7}, and references therein. In \cite{intr11} the well--known passivity property of \cite{SANVER} is used to address an ``approximate" tracking problem for an
inverter connected to a photovoltaic solar panel. A similar framework was studied in \cite {meza2}.

The remainder of this paper is organized as follows. Problem formulation is presented in Section 2. Our main theoretical result is contained in Section 3, where a linear matrix inequality (LMI) condition is
imposed to solve the tracking problem---invoking passivity theory. Section 4 is devoted to the synthesis of a PI controller that ensures tracking trajectory, under some suitable detectability assumptions. The
result is applied in Sections 5 and 6 to the two power electronic applications mentioned above. Simulations and experimental results are included in these two sections. Finally, conclusions in Section 7 complete the paper. A preliminary version of this paper was reported in \cite{tracking}
%
\section{Global Tracking Problem}
%
Consider the bilinear system
\begin{equation}
\dot{x}(t)=Ax(t)+d\left(t\right)+\sum^{m}_{i=1}u_i(t)B_ix(t)
\label{eq:bilinear}
\end{equation}
where\footnote{For brevity, in the sequel the time argument is omitted from all signals.} $x\in\mathbb{R}^{n},d\in\mathbb{R}^{n}$ are the state and the (measurable) disturbance vector, respectively, $u \in
\mathbb{R}^m,\; m \leq n$, is the control vector, and $A \in \mathbb{R}^{n\times n},B_i \in \mathbb{R}^{n\times n}$ are real constant matrices.

We will say that a function $x_\star:\mathbb{R}_+ \to \mathbb{R}^n$ is an {\em admissible trajectory} of \eqref{eq:bilinear}, if it is differentiable, bounded and  verifies
\begin{align}
\dot{x}_{\star}=&Ax_\star+d+\sum_{i=1}^{m}u_i^\star B_ix_\star
\label{xsta}
\end{align}
for some bounded control signal $u_\star:\mathbb{R}_+ \to \mathbb{R}^m$.

The global tracking problem is to find, if possible, a dynamic state--feedback controller of the form
\begin{eqnarray}
\dot z & = & F(x,x_\star,u_\star )\\
u & = & H(x,x_\star,u_\star ),
\end{eqnarray}
where $F:\mathbb{R}^n \times \mathbb{R}^n \times \mathbb{R}^m \to \mathbb{R}^q,\;q \in {\mathbb Z}_+$, and $H:\mathbb{R}^n \times \mathbb{R}^n \times \mathbb{R}^m \to \mathbb{R}^m$, such that all signals
remain bounded and
\begin{equation}
\label{glotra}
\lim_{t \to \infty} [x(t)-x_\star (t)]=0,
\end{equation}
for all initial conditions $(x(0),z(0)) \in \mathbb{R}^n \times \mathbb{R}^q$ and all admissible trajectories.

In this paper a set of matrices $\{A,B_i\}$ has been characterized for which it is possible to solve the global tracking problem with a simple {\em linear time--varying PI controller}. The class is identified via the
following LMI.

\begin{assumption}\em
\label{ass1}
$\exists P \in \mathbb{R}^{n \times n}$ such that
\begin{align}
P=P^\top >0 \\
\mbox{sym}(PA)\leq 0
\label{pa}\\
\mbox{sym}(PB_i)=0,
\label{pbi}
\end{align}
where the operator $\mbox{sym}:\mathbb{R}^{n \times n} \to \mathbb{R}^{n \times n}$ computes the symmetric part of the matrix, that is
$$
\mbox{sym}(PA)=\hal(PA+A^\top P).
$$
\end{assumption}

To simplify the notation in the sequel the {\em positive semidefinite} matrix has been defined
\begin{equation}
\lab{q}
Q:=-\mbox{sym}(PA).
\end{equation}
%
\section{Passivity of the Bilinear Incremental Model }
%
Instrumental to establish the main result of the paper is the following lemma.

\begin{lemma}
\label{lem1}
Consider the system \eqref{eq:bilinear} verifying the LMI of Assumption \ref{ass1} and an admissible trajectory $x_\star$. Define the incremental signals
$$
\tilde {(\cdot)}:= (\cdot)- (\cdot)_\star ,
$$
and the $m$--dimensional output function
\begin{equation}
y:=\mathcal{C}(x_\star){x}
\label{eq:ypassive}
\end{equation}
where the map $\mathcal{C}:\mathbb{R}^n \to \mathbb{R}^{m\times n}$ is defined as
\begequ
\lab{c}
\mathcal{C}(x_\star):=\left[\begin{matrix}x_\star^\top B_1^\top \\\vdots\\x_\star^\top B_m^\top \end{matrix}\right]P.
\endequ
The operator $\tilde{u}\mapsto y$ is {\em passive} with storage function
\begin{equation}
V(\tilde{x}):=\frac{1}{2}\tilde{x}^\top P\tilde{x}.
\label{eq:V}
\end{equation}
Hence, it verifies the dissipation inequality
\begin{equation*}
\dot{V}\leq \tilde{u}^\top y.
\end{equation*}
\end{lemma}
\begin{pf}
Combining \eqref{eq:bilinear} and \eqref{xsta} yields
\begin{align}
\dot{\tilde{x}}=& (A+\sum_{i=1}^mu_iB_i)\tilde{x}+\sum_{i=1}^m\tilde{u}_iB_ix_\star.
\label{eq:modelinc}
\end{align}
Now, the time derivative of the storage function \eqref{eq:V} along the trajectories of \eqref{eq:modelinc}  is
\begin{eqnarray*}
\dot{V}(\tilde{x}) & = & \tilde{x}^\top P\left[ (A+\sum_{i=1}^mu_iB_i)\tilde{x}+\sum_{i=1}^m\tilde{u}_iB_ix_\star\right]\\
                  & = & -\tilde{x}^\top Q\tilde{x}+\sum_{i=1}^m\tilde{u}_i\tilde{x}^\top PB_ix_\star\\
                   & \leq & \sum_{i=1}^m\tilde{u}_i\tilde{x}^\top PB_ix_\star\\
                   & = & \sum_{i=1}^m\tilde{u}_i{x}^\top PB_ix_\star\\
                   & = &  y^\top \tilde{u},
\end{eqnarray*}
where \eqref{pbi} of Assumption \ref{ass1} has been used to get the second identity, \eqref{pa} for the first inequality, \eqref{pbi} again for the third equation and  \eqref{eq:ypassive} for the last
identity.
\end{pf}

\begin{remark}
\label{rem0}
A key step for the utilization of the previous result is the derivation of the desired trajectories $x_\star$ and their corresponding control signals $u_\star$, which satisfy \eqref{xsta}. As shown in the
examples below this may prove to be a very complicated task and some approximations may be needed to derive them. Indeed, it is shown in \cite{intr5} that even for the simple boost converter this task involves the search of a stable  solution of an Abel ordinary differential equation, which is known to be highly sensitive to initial conditions.
\end{remark}
%
\section{A PI Global Tracking Controller}
%
From Lemma \ref{lem1} the next proposition follows immediately.
\begin{proposition}
Consider the system \eqref{eq:bilinear} verifying Assumption \ref{ass1} and an admissible trajectory $x_\star$ in closed loop with the PI controller
\begin{align}
\nonumber
\dot{z}=&-y\\
u=&-K_py+K_iz+{u}_\star\label{eq:PIcont}
\end{align}
with output \eqref{eq:ypassive}, \eqref{c} and $K_p=K_p^\top >0$, $K_i=K_i^\top >0$. For all initial conditions $(x(0),z(0))\in \mathbb{R}^{n} \times \mathbb{R}^{m}$ the trajectories of the closed-loop system
are bounded and
\begin{equation}
\label{ya}
\lim_{t\rightarrow\infty}y_a(t)=0,
\end{equation}
where the augmented output $y_a:\mathbb{R}_+ \to \mathbb{R}^{m+n}$ is defined as
\begin{equation*}
y_a:=\begin{bmatrix}\mathcal{C}(x_\star)\\ Q^\hal \end{bmatrix}\tilde{x},
\end{equation*}
with $Q^\hal$ the square root of $Q$ given in \eqref{q}. Moreover, if
\begin{equation}
\label{ran}
\mbox{rank}\begin{bmatrix}\mathcal{C}(x_\star)\\ Q^\hal\end{bmatrix} = n,
\end{equation}
then state global tracking is achieved, {\em i.e.}, \eqref{glotra} holds.
\end{proposition}
\begin{pf}
Notice that the PI controller \eqref{eq:PIcont} is equivalent to
\begin{align*}
\tilde{u}=& \nonumber-K_py+K_i{z}\\
\dot {z}=&-y.
\end{align*}
Propose the following radially unbounded Lyapunov  function candidate
\begin{equation*}
W(\tilde x,z):=V(\tilde{x})+\frac{1}{2}{z}^\top K_i{z},
\end{equation*}
whose time derivative is
\begin{eqnarray*}
\dot{W}&=&-\tilde{x}^\top Q\tilde{x}+y^\top \tilde{u}-{z}^\top K_i{y}\label{eq:der1}\\
&=&-\tilde{x}^\top Q\tilde{x}-y^\top K_py \\
& \leq &-\lambda_{\min}\{K_p\}|y|^2-|Q^\hal\tilde{x}|^2 \leq 0\label{eq:der2}
\end{eqnarray*}
From here we conclude that the system state $z,\tilde x \in \Linf$ and $y, Q^\hal\tilde{x} \in \Lt$, consequently $y_a \in \Lt$. To conclude that $y_a(t) \to 0$ it suffices, invoking \cite{tao}, to prove that
$\dot y_a \in \Linf$. Towards this end, we first notice that $\tilde x,x_\star \in \Linf$ implies $x \in \Linf$ and, this in its turn, implies from \eqref{eq:ypassive} $y \in \Linf$. Now,  $y,z,u_\star \in
\Linf$ implies, from \eqref{eq:PIcont}, $u \in \Linf$. That implies, from \eqref{eq:modelinc}, $\dot{\tilde x} \in \Linf$. Now, compute
\begin{equation}
\dot y=\begin{bmatrix}\dot x_\star^\top B_1^\top\\\vdots\\\dot x_\star^\top B_m^\top\end{bmatrix}P\tilde x+\begin{bmatrix}x_\star^\top B_1^\top\\\vdots\\x^\top_\star B_m^\top\end{bmatrix}P\dot{\tilde x},
\end{equation}
which is bounded because $\dot x_\star \in \Linf$.

The proof of global state tracking follows noting that  $y_a(t) \to 0$ ensures \eqref{glotra} if the rank condition \eqref{ran} holds.
\end{pf}

\begin{remark}
\label{rem1}
Notice that the matrix $\mathcal{C}$ depends on the reference trajectory. Therefore, the rank condition \eqref{ran} identifies a class of trajectories for which global tracking is ensured.
\end{remark}

\begin{remark}
Condition \eqref{ran} is sufficient, but not necessary for state convergence. Indeed, as shown in \cite{schaft}, global tracking is guaranteed if $y_a$ is a detectable output for the closed--loop system. That
is, if the following implication holds
\begin{equation*}
y_a(t)\equiv 0\implies\lim_{t\rightarrow\infty}[x(t)-x_\star(t)]=0.
\end{equation*}
\end{remark}
%
%
\section{Application to Power Factor Compensation in a Boost Converter}
%
%
\subsection{ Model and Problem Formulation}

The following scenario corresponds to the PFC of an AC--DC boost converter. Assuming linear loads and sinusoidal steady--state regime it is well--known that the power factor is optimized when the line current is
in phase with line voltage.\footnote{See \cite{OrtegaNonLin} for the case of nonlinear loads and non--sinusoidal regime.} Hence, the problem can be recast as tracking problem that fits the theoretical
framework given in the previous sections. An additional requirement imposed to the PFC is to reduce the distortion in the current, which can also distort the line voltage. A standard measure to assess this
property is the Total Harmonic Distortion (THD) index.

\begin{figure}[t]
\centering
\includegraphics{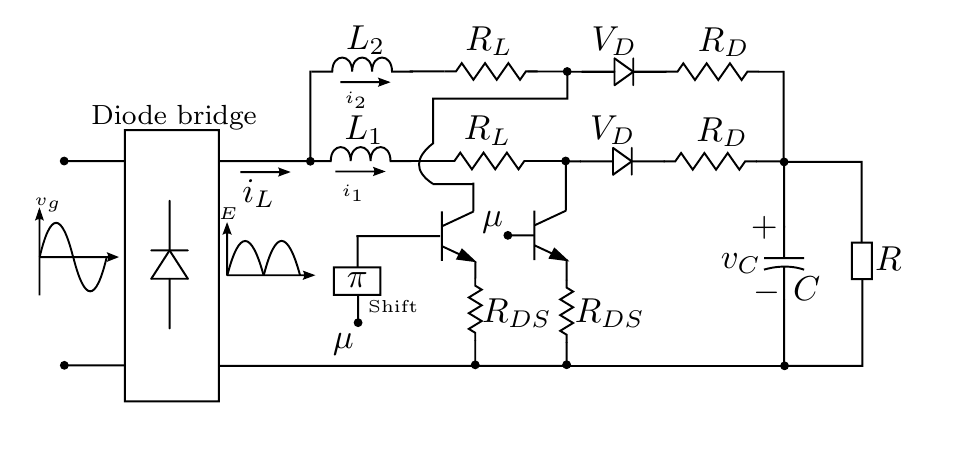}
\caption{Boost converter circuit}
\label{fig:boost}
\end{figure}

The interleaved  AC--DC boost is one of the most popular PFC topologies in real applications. The idea of this topology is to incorporate $N$ branches with their own switch, each of whom admitting the control signal shifted by $\frac{2\pi}{N}$. When assuming parasitic parameters such as the voltage drop or losses in the diodes, a typical form of a two branches converter is depicted in Fig. \ref{fig:boost}. Since these parameters are usually small, they will be neglected. Hence, from Fig. \ref{fig:boost},  $R_{DS}, R_D, R_L=0 $ .  Under this consideration, the model equations are 

\begin{align}
\dot i_{1}=&-\frac{1}{L_1}(1-\mu')v_C+\frac{E}{L_1} \nonumber\\
\dot i_{2}=&-\frac{1}{L_2}(1-\mu')v_C+\frac{E}{L_2} \nonumber \\
\dot v_C=&\frac{1}{C}(1-\mu') i_{1} + \frac{1}{C} (1-\mu') i_{2} -\frac{1}{RC}v_C, \label{eq:interN}
\end{align}

\noindent where $E$ is the rectified AC voltage $L_1, L_2,C,R$ are the boost inductance, capacitance and load resistance, respectively, and $\mu=1-\mu'$ is the duty cycle of the switch . 
If it is assumed that $L_1=L_2=\frac{L}{2}$, it follows that $i_{1}=i_{2}=\frac{1}{2}i_L$ and, according to Kirchhoff's law, $i_L =i_{1} + i_{2} $ (see Fig.  \ref{fig:boost}). Then, using the latter and  defining $x:=\begin{bmatrix}i_L&v_C\end{bmatrix}^\top$ and $u = 1-\mu'$, equations \eqref{eq:interN} can be expressed as 

\begin{align}
\dot{x}_1=& -\frac{2}{L} u x_2+\frac{2}{L}E \nonumber \\
\dot{x}_2=&\frac{1}{C} u x_1-\frac{1}{RC}x_2,  \label{eq:mboost}
\end{align}

Clearly, the system \eqref{eq:mboost} can be written in the form \eqref{eq:bilinear} with the following definitions
\begin{align*}
A &= \left[
\begin{matrix}
0 & 0 \\ 0 & -\frac{1}{RC}
\end{matrix}\right],
d=\left[
\begin{matrix}
\frac{2E}{L}\\ 0
\end{matrix}\right],B=\left[
\begin{matrix}
0 & -\frac{2}{L}\\\frac{1}{C} & 0
\end{matrix}
\right].
\end{align*}


%
%
A matrix $P$ that verifies the conditions of Assumption \ref{ass1} is
\begin{equation}
P = \left[
\begin{matrix}
\frac{L}{2} & 0 \\ 0 & C
\end{matrix}
\right],
\end{equation}

\noindent then

\begin{equation}
Q = \left[
\begin{matrix}
0 & 0 \\ 0 & \frac{1}{R}
\end{matrix}
\right].
\end{equation}
Furthermore, the passive output $y$  is
\begin{align}
y=&x_\star^\top B^\top x \nonumber\\
=&x_{1\star} x_2-x_{2\star}x_1\label{eq:output2}.
\end{align}
Now, convergence of the state is tested by means of the matrix
$$
\begin{bmatrix}\mathcal{C}(x_{\star})\\ Q^\hal \end{bmatrix}=
\begin{bmatrix}-x_{2\star}&x_{1\star}\\0&0\\0&\frac{1}{\sqrt{R}}\end{bmatrix},
$$
which satisfies the  rank condition \eqref{ran}, provided $x_2^{\star}$ is bounded away from zero, which is consistent with the operating mode of the boost, i.e., $x_2>0$.

\noindent Reference signals $x_{2\star}$, $x_{1\star}$ and  $u_{\star}$ are derived below.
\subsection{Generation of references}
It is assumed that the dynamics of the output voltage $x_2$ is slower respect to that of the current $x_1$. 
The objective of control is to drive $x$ to $x_{\star}$ in such a way that
\begin{enumerate}
\item $x_{1\star}\propto E$ (Power Factor Correction)
\item $x_{1\star} \propto \frac{1}{E_{RMS}} $.
\item $x_{2\star} $ is a desired output voltage.
\end{enumerate} 

Then, the resulting reference for the inductor current is shown in the following equation:
\begin{equation}
x_{1 \star} = \frac{E \phi}{(E_{RMS})^2}
\end{equation}
where $\phi$ is the output of a linear compensator $G_{cv}(s)$ whose input is the voltage error and $E_{RMS}$ is the RMS value of the rectified AC voltage. The design of the compensator $G_{cv}(s)$ is based on the perturbation and linearization technique applied in the
loss-free resistor (LFR) boost model \cite{erickson2001fundamentals}. This results in equivalent small-signal circuit for the design of the following transfer function \cite{VoltageLoopDesign}
\begin{equation}
\frac{\hat x_2(s)}{\hat \phi (s)}=\frac{P_{av}}{sCX_2 \Phi},
\end{equation}
where $\hat x_2(s)$ and $X_2$ are respectively the Laplace small signal of $x_2$ and its DC output voltage, also $\hat \phi(s)$ and $\Phi$ are respectively the Laplace small signal of $\phi$ and its DC
compensator output control signal, $P_{av}$ is the average rectifier power and $C$ is the boost capacitance. This transfer function neglects the complicating effects of high-frequency switching ripple, and is
valid for control variations at frequencies sufficiently lower than the AC line frequency. Ultimately, the linear compensator $G_{cv}(s)$ consists of a PI with anti-windup technique
\cite{erickson2001fundamentals}. The output voltage controller must have sufficiently small gain at frequency and minimize the negative effect of the Right Half Plane (RHP) zero. Hence its bandwidth must be
low. As a rule of thumb, setting the overall control loop bandwidth to a third of the RHP zero is enough to provide the closed loop stability.  It requires, however, a compromise in the control performance.
\cite{erickson2001fundamentals}.

The design of the controller is concluded with the calculation of   $u_{\star}$. Thus, from the first equation of \eqref{eq:mboost}
\begin{equation}
u_\star=\frac{2E-L\dot x_{1\star}}{2x_{2\star}}\label{eq:cboost}.
\end{equation}
\subsection{The Controller}
%
\begin{figure}[t]
\centering
\includegraphics[scale=1.3]{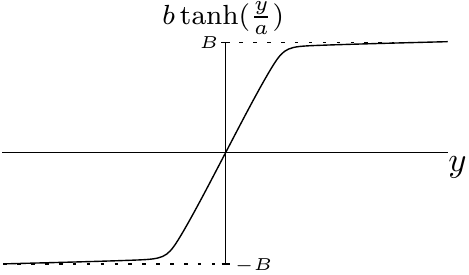}
\caption{Function $B\tanh(\frac{y}{A})$}
\label{fig:tanh}
\end{figure}
In summary, the resulting controller equations for the PFC AC-DC Boost Converter  are the following.

\begin{align}
\dot z&=-y\nonumber\\
u&=-K_p y+K_i z+ u_{\star}\nonumber \\
y   &= x_{1\star} x_2-x_{2\star}x_1\nonumber\\
u_\star & =\frac{2E-L\dot x_{1\star}}{2x_{2\star}}\nonumber\\
x_{1 \star} &= \frac{E \phi}{(E_{RMS})^2} \label{eq:finconBC}
\end{align}
where $\phi$ is the output of the anti-windup PI and $x_{2\star}$ is the desired DC output voltage.
\begin{remark}
It is customary to implement a PI controller with changing gains. Usually nonlinear functions as the $\tanh(\cdot)$ are introduced. In this case, a modification in the second equation  of \eqref{eq:finconBC}
leads to
\begin{equation}\label{tanhu}
u=-b\tanh\left(\frac{y}{a}\right)+K_i z + u_{\star}
\end{equation}
where $a$ and $b$ are free parameters to be adjusted. A typical shape of function $b\tanh(\cdot)$ is depicted in  Fig. \ref{fig:tanh}. As a matter of fact this change does not compromise the stability of the
system. It comes down to the fact that the function $\tanh(y)$ can be considered as the product $k_{p}(y)y$ where $k_p(y)$ is scalar function taking nonnegative bounded values except at $y=0$ where it is zero. A practical comparison between controller
\eqref{eq:finconBC} and its modification will be shown in the sequel.
\end{remark}
\subsection{Simulations and Experimental Results}
The interleaved Boost converter was simulated under the following considerations: the output voltage reference $Vref=15V$, the  PWM frequency $PWM_{Freq}=100KHz$ and the input voltage $v_g=9|sin(2 \pi f_{line} t)|$ where $f_{line}=50Hz$.

\begin{table}[t]
	\begin{center}
		\caption{}\label{tab:indexes}
		\begin{tabular}{l|cc}
			Indexes & PF & THD \\
			\hline
			$Proportional$ & $96.7\%$ & $23.2\%$  \\
			$Hyperbolic$ $Tangent$ & $98.2\%$  & $21.1\%$  \\
		\end{tabular}
	\end{center}
\end{table}

Boost inductance, capacitance and load values are respectively: $L=56\mu H$, $C=3047\mu F$, $R=22\Omega$. 
The gain for the compensator $G_{cv}(s)$ and for the passivity PI have been respectively set to $k_{p(comp)}=0.011$, $k_{i(comp)}=0.03$ and $K_p=0.013$, $K_i=0.0001$ whereas the hyperbolic tangent parameters
are $a=55$ and $b=0.25$.

\begin{figure} [t]
 \captionsetup[subfigure]{labelformat=empty}
 \centering
 \subfloat[Simulation Trend  ]{\includegraphics[width=0.48\linewidth]{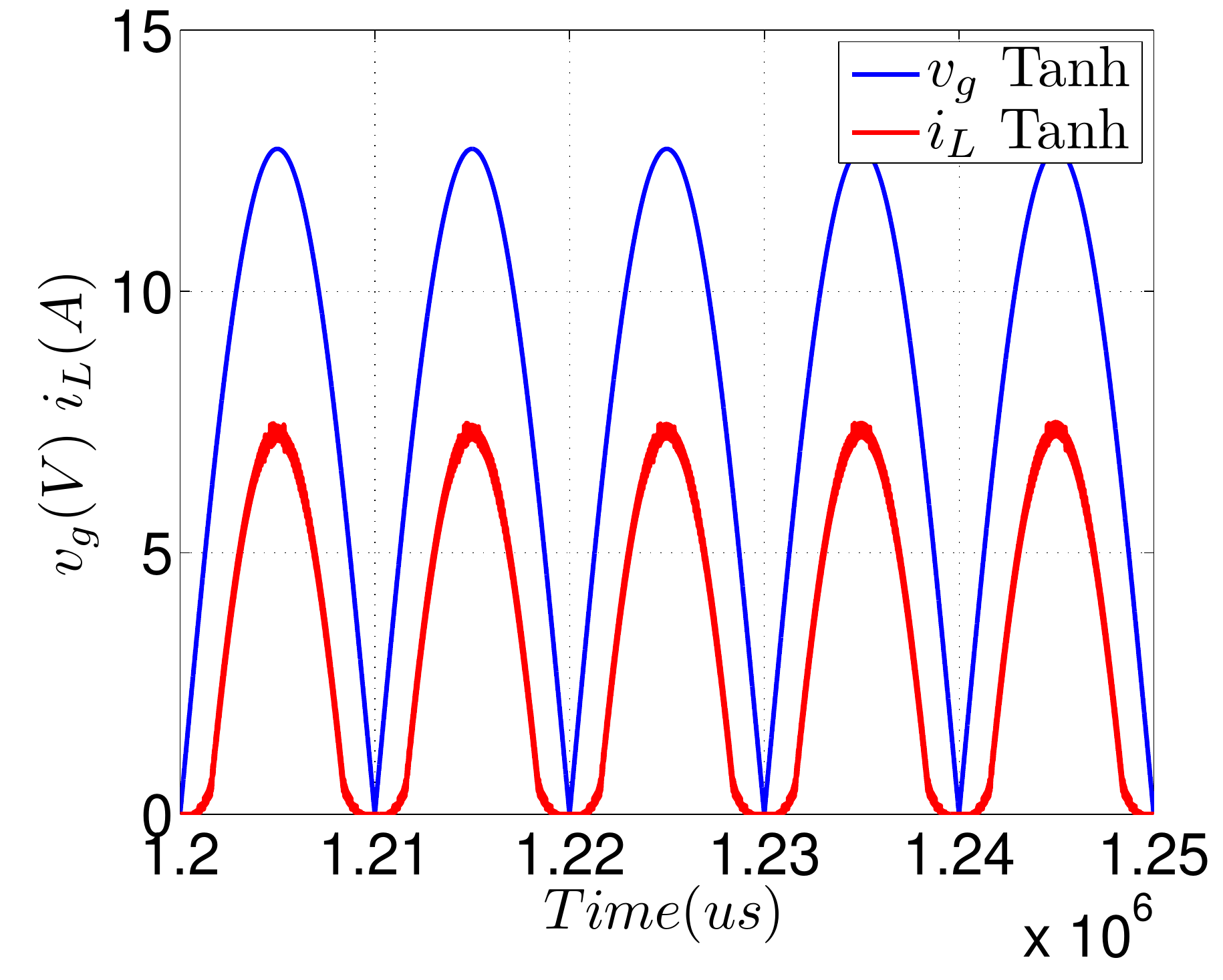}}\quad
 \subfloat[Experimental Trend.]{\includegraphics[width=0.48\linewidth]{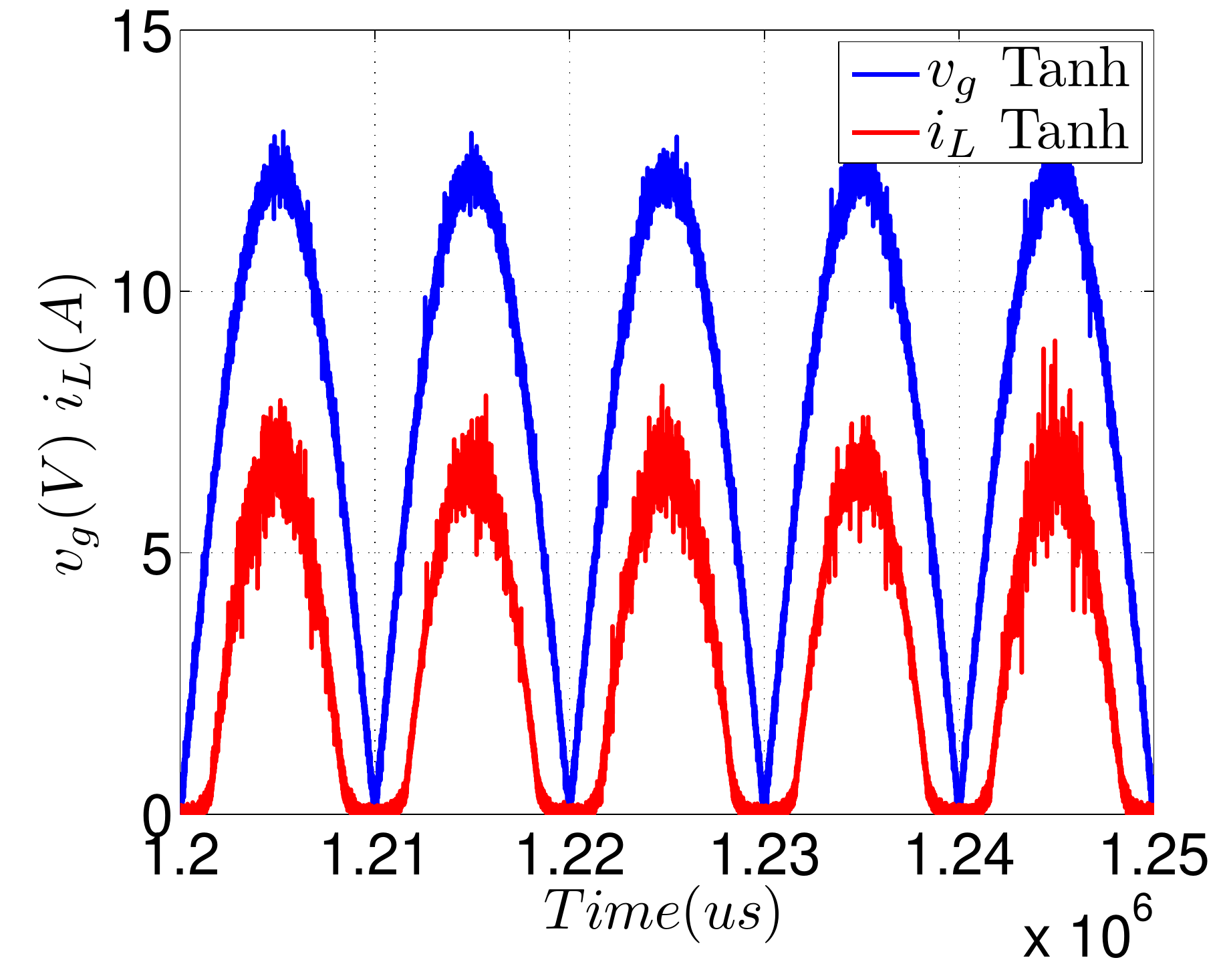}}
 \caption{Inductor Current $i_L$ and Input Voltage $v_g$ Trend.} \label{fig:i1}
 \subfloat[Simulation Trend.]{\includegraphics[width=0.48\linewidth]{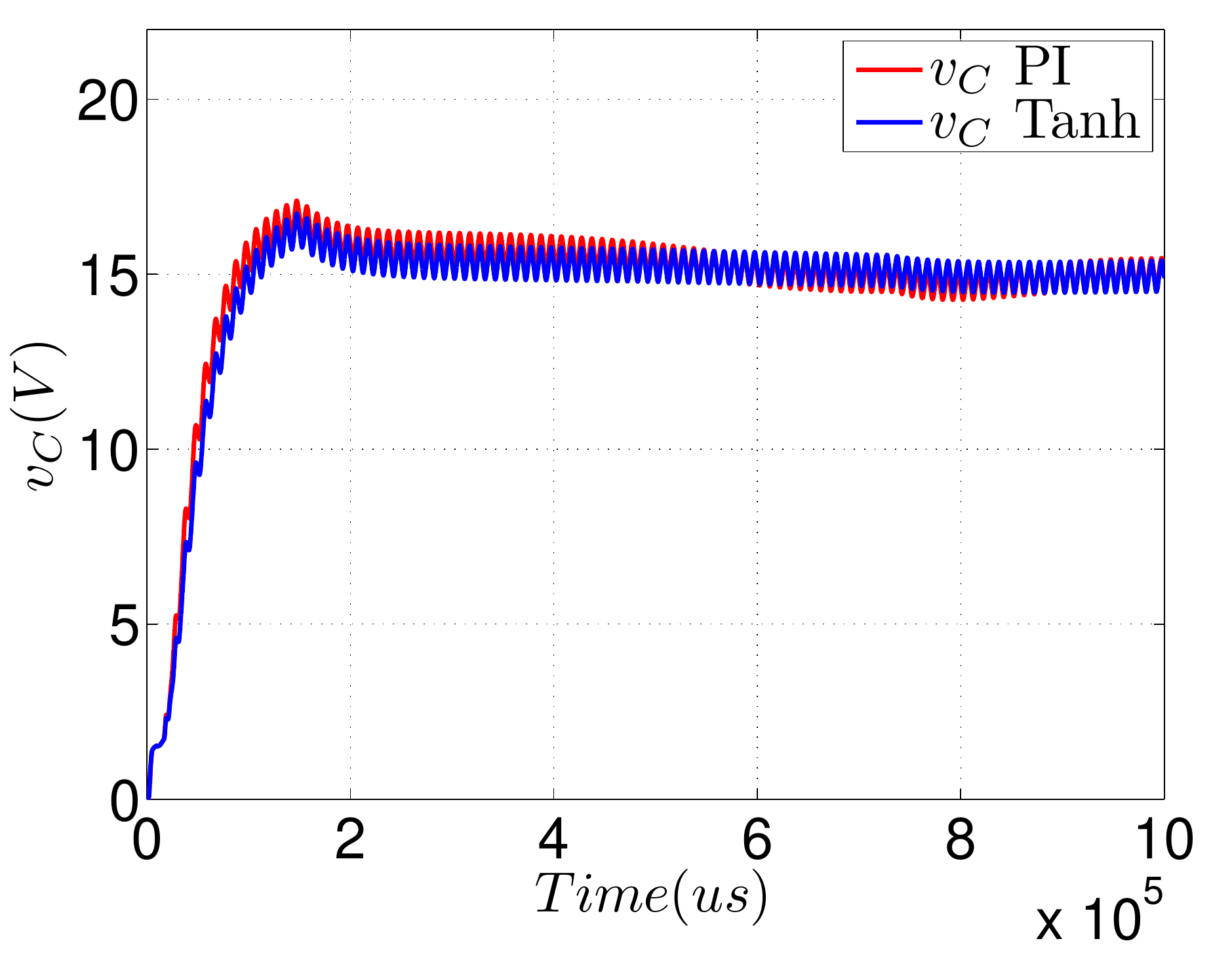}}\quad
 \subfloat[Experimental Trend.]{\includegraphics[width=0.48\linewidth]{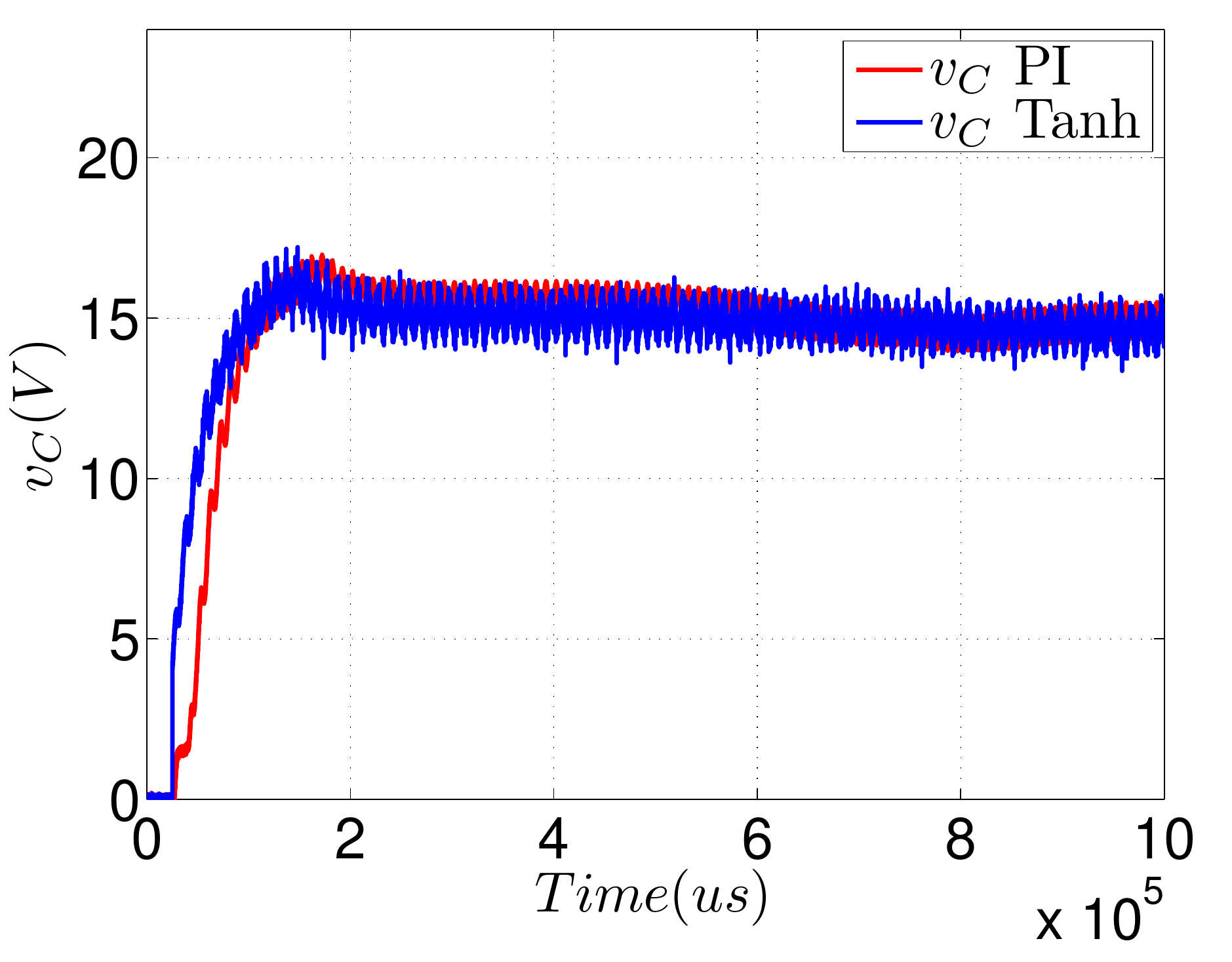}}
  \caption{Output Voltage $v_C$ Transient. Comparison between \eqref{eq:finconBC} and \eqref{tanhu}}\label{fig:vc}
  \subfloat[Simulation Trend.]{\includegraphics[width=0.48\linewidth]{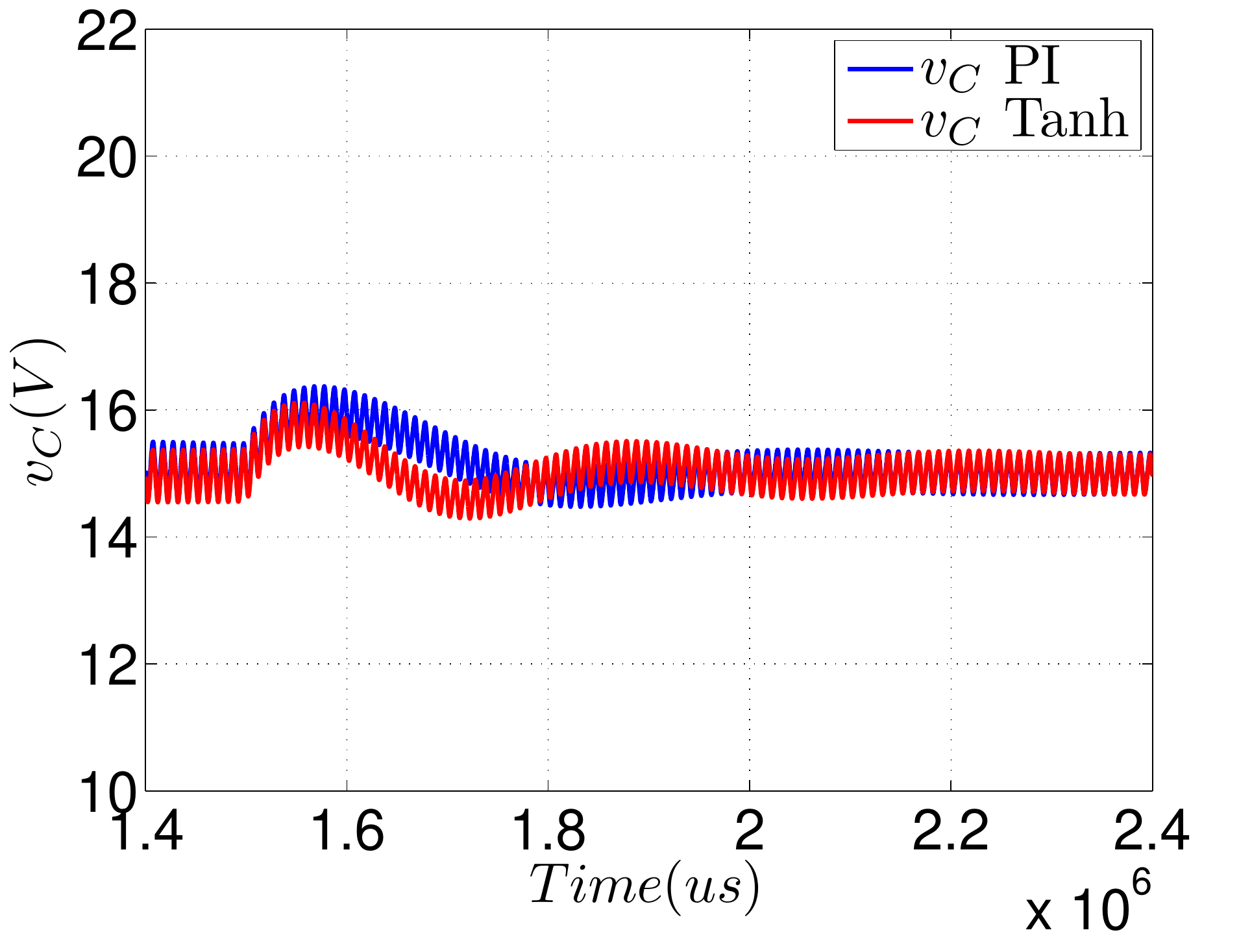}}\quad
  \subfloat[Experimental Trend.]{\includegraphics[width=0.48\linewidth]{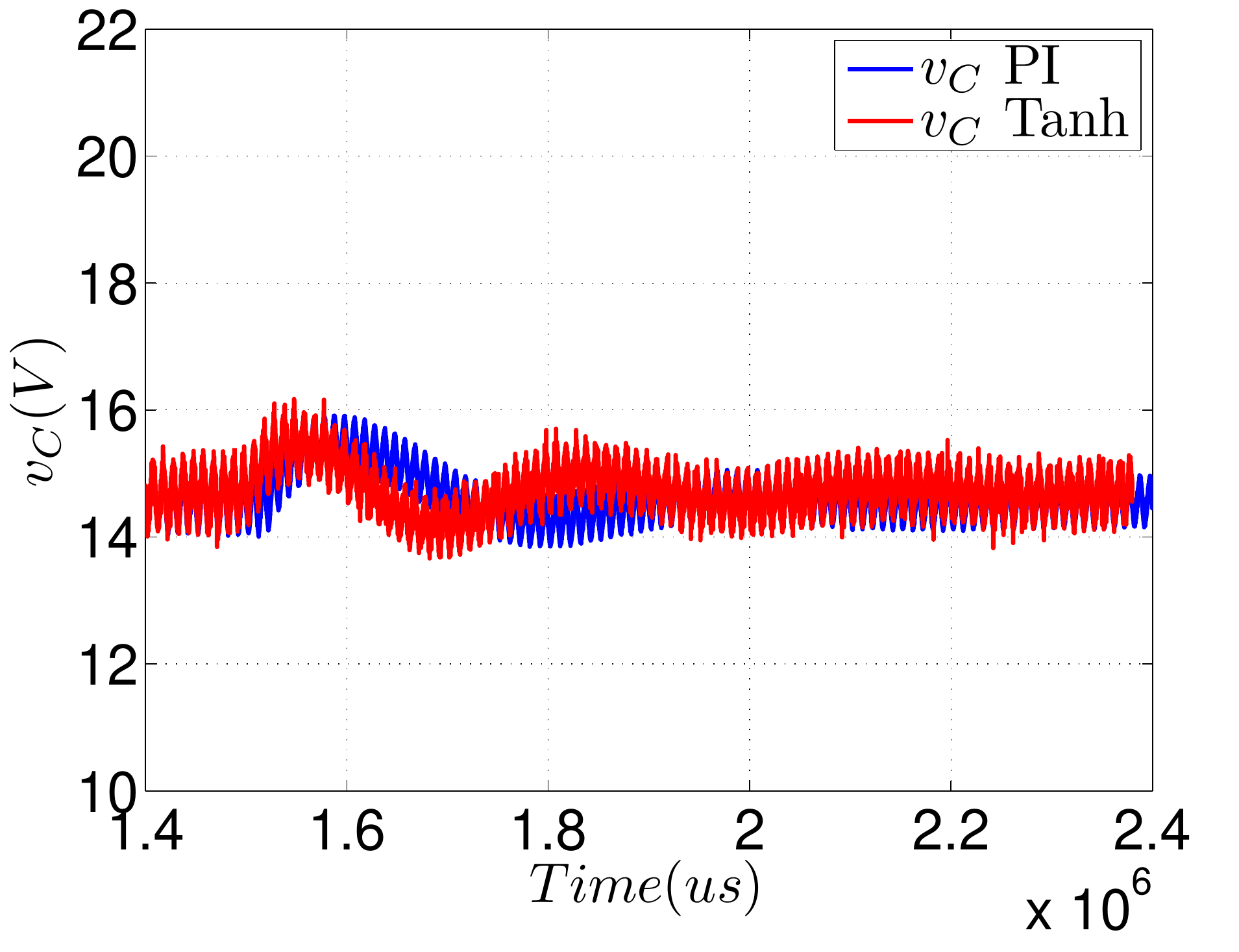}}
  \caption{Output Voltage $v_C$ Trend with a load change of $30\%$.  Comparison between \eqref{eq:finconBC} and \eqref{tanhu}} \label{fig:load_change}
\end{figure}

Figure \ref{fig:i1} shows the trend of the inductor current $i_L$ and of the input voltage $v_g$ in an interval time during the steady state: it is clear that the current $i_L$ (the blue line) and input
voltage $v_g$ (the red line) have the same shape, except in the instants in which the value of $v_g$ is low driving the boost to the Discontinuous Conduction Mode (DCM). Also performance indexes, displayed in
Table \ref{tab:indexes}, confirm that the control reaches the main objective and suggest the use of the hyperbolic tangent instead of PI because of its slightly better performances. Figure \ref{fig:vc} shows
the trends of the output voltage $v_C$. Its reference value, namely $15V$, is completely reached in $0.8s$. Finally, Figure \ref{fig:load_change} shows $v_C$ trend after a load change of $30\%$ which occurs at
$1.5s$. As it can be seen, after an initial transient the system is stabilized again around the reference value. Even if the gains can be tuned to slightly change the performances achieved, in general it could be
stated that the hyperbolic tangent function has better performances in the steady state, both as regards the disturbance response and the performance indicators (PF and THD).

%
\section{Phase Independent Control for Modular Multilevel Converter}
%
%
\subsection{ Model and Problem Formulation}
The MMC introduced by Prof. Marquart in \cite{MMC} has several advantages with respect to its predecessors, such as its high modularity, scalability and lower losses. It seems to be one of the most promising
converters for bulk power transmission via HVDC links \cite{Pros}.

Nonetheless, controlling the MMC is no simple task.  Several efforts have been oriented to propose suitable mathematical dynamical models \cite{Ondyn,Dyn}, and control strategies for the converter under
balanced operation. Some of the proposed control strategies have been implemented in $dqo$ rotating reference frame such as \cite{CCSC,gbJour}. Nonetheless, it seems to be getting clear that there are
significant disadvantages in applying $dqo$--based control schemes for the control of the MMC differential currents since they do not facilitate complete phase--independent control of the converter state
variables, hence the MMC potential will not be fully exploited \cite{gbJour2}. To overcome this drawback it is necessary to formulate the control problem in the $abc$ frame---in which the MMC is controlled
independently per phase---resulting in a tracking problem instead of a regulation one.

\begin{figure}
\includegraphics[scale=0.9]{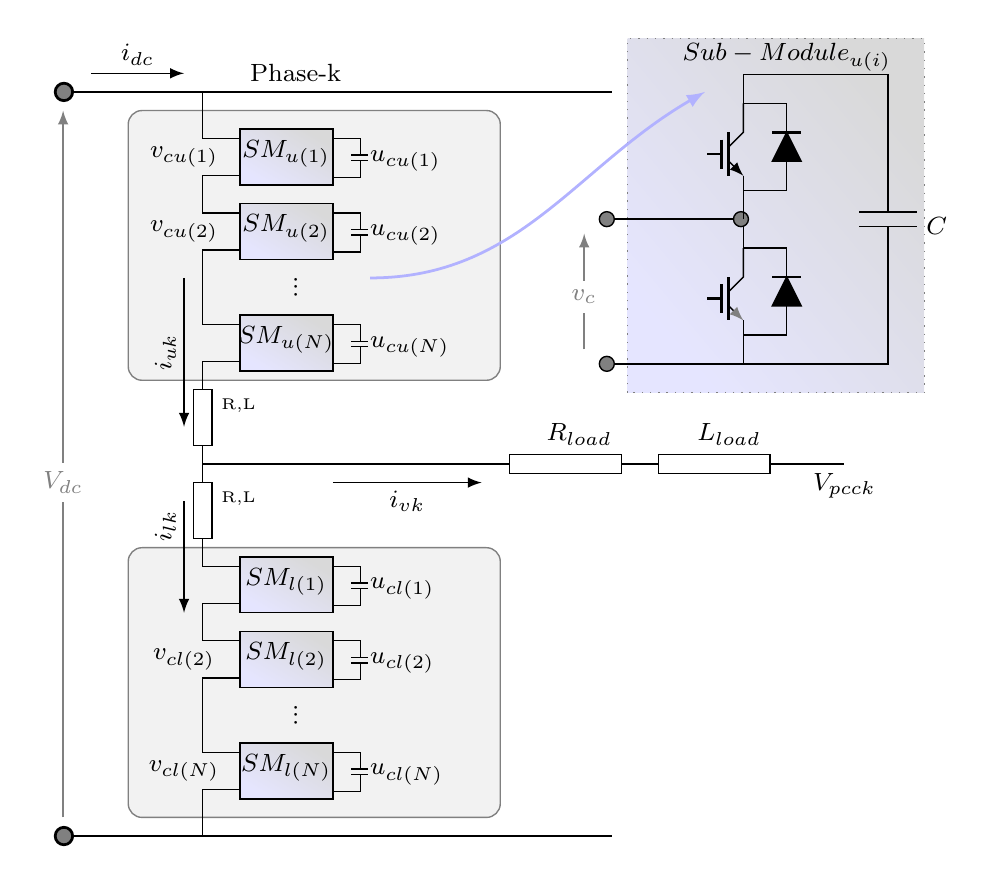}
\caption{MMC Topology}
\label{fig:vsc}
\end{figure}

The MMC converter studied in the paper is shown in Fig. \ref{fig:vsc}. To obtain its mathematical model four state variables have been considered: 1) The differential or circulating current of the MMC $i_{diff}$, 2)
the grid or load current $i_{v}$, 3) the sum between the upper and lower capacitor voltages $u_{C\Sigma}:=u_{CU}+u_{CL}$, and 4) the difference between them $u_{C\Delta}:=u_{CU}-u_{CL}$. This yields the model
\begin{align}
\dot i_{diff} =& - \frac{R}{L} i_{diff}-\frac{1}{4L} {u}_{C\Sigma}u_\Sigma - \frac{1}{4L}  {u}_{C\Delta}u_\Delta+\frac{v_{dc}}{2L} \nonumber\\
\dot{i}_v =&- \frac{R'}{L'}{i}_v  - \frac{1}{4L'} {u}_{C\Delta}u_\Sigma-\frac{1}{4L'} {u}_{C\Sigma}u_\Delta - \frac{V_{pcc}}{L'}\nonumber\\
\dot{u}_{C\Sigma} =&\frac{1}{C'}  i_{diff} u_\Sigma +  \frac{1}{2C'}{i}_vu_\Delta\nonumber\\
\dot{u}_{C\Delta} =&   \frac{1}{2C'}{i}_vu_\Sigma + \frac{1}{C'}i_{diff}u_\Delta,  \label{eq:mmcmodel}
\end{align}
where the control signals are $u_\Sigma$ and $u_\Delta$ and the other signals and constants can be identified from Fig. \ref{fig:vsc} (Refer to \cite{Dyn,Ondyn} for more details on the model). Also, henceforth $Vpcc=0$ since the system here studied is not considered to be connected to grid. In addition, notation for the equivalent capacitance $C':=C/N$, the equivalent inductance $L':=L/2 + L_{load}$ and resistance $R':=R/2 + R_{load}$ has been introduced for the sake of clarity. Defining $x: =\begin{bmatrix}i_{diff} & i_{v} & u_{C\Sigma}  & u_{C\Delta}\end{bmatrix}^\top$ and $u: =\begin{bmatrix}u_{\Sigma}  & u_{\Delta}\end{bmatrix}^\top$, the system can be written in the form \eqref{eq:bilinear} with:
\begin{align*}
\nonumber A =& \left[
\begin{matrix}
-\frac{R}{L}&0&0&0\\0&-\frac{R'}{L'}&0&0\\0&0&0&0\\0&0&0&0
\end{matrix}
\right],d=\left[
\begin{matrix}
\frac{v_{dc}}{2L}\\ 0 \\0\\0
\end{matrix}
\right],\\ \nonumber B_1 =& \left[
\begin{matrix}
0&0&\frac{-1}{4L}&0\\0&0&0&\frac{-1}{4L'}\\ \frac{1}{C'}&0&0&0\\ 0& \frac{1}{2C'} &0&0
\end{matrix}
\right], B_2 = \left[
\begin{matrix}
0&0&0&\frac{-1}{4L}\\ 0&0&\frac{-1}{4L'}&0 \\ 0&\frac{1}{2C'}&0&0\\ \frac{1}{C'}&0&0&0
\end{matrix}
\right].
\end{align*}
Therefore, the matrix $P$ satisfying  Assumption \eqref{ass1} is
$$
P=\begin{bmatrix}2L&0&0&0\\0&L'&0&0\\0&0&\frac{C'}{2}&0\\0&0&0&\frac{C'}{2}\end{bmatrix}.
$$
The passive output defined in \eqref{eq:ypassive} is
\begin{align}
y=& \left[
\begin{matrix}
x_{\star}^\top B_1^\top \\ x_{\star}^\top B_2^\top
\end{matrix}
\right]Px\nonumber\\
=&\frac{1}{2}\left[
\begin{matrix}
x_{1\star} x_{3}-x_{1}x_{3\star}+\frac{1}{2}x_{2\star}x_{4}-\frac{1}{2}x_2x_{4\star}\\
x_{1\star}x_{4}-x_{1}x_{4\star}+\frac{1}{2}x_{2\star}x_{3}-\frac{1}{2}x_2x_{3\star}
\end{matrix}
\right] \label{eq:output1}.
\end{align}
The convergence condition \eqref{ran} imposes that the rank of the matrix
$$
\begin{bmatrix}\mathcal{C}\\ Q^{\frac{1}{2}}\end{bmatrix}=
\begin{bmatrix} -\frac{1}{2}x_{3\star}&-\frac{1}{4}x_{4\star}&\frac{1}{2}x_{1\star}&\frac{1}{4}x_{2\star}\\-\frac{1}{2}x_{4\star}&-\frac{1}{4}x_{3\star}&\frac{1}{4}x_{2\star}&\frac{1}{2}x_{1\star}\\\sqrt{2R}&0&0&0\\0&\sqrt{R'}&0&0\\0&0&0&0\\0&0&0&0\end{bmatrix}
$$
should be full. This is clearly the case if
\begin{equation} \label{rankc}
x_{1\star}^2-{1 \over 4}x_{2\star}^2 \neq 0,
\end{equation}
%
\subsection{Generation of references}
As indicated in Remark \ref{rem0} obtaining the exact expressions for the desired trajectories $x_\star$ and their corresponding control signals $u_\star$ that satisfy \eqref{xsta}  is a very complicated task.
Therefore, in this subsection several practical considerations have been made, which are widely adopted in the power electronics community, to approximate their solution.

The reference equilibrium point for the grid current is imposed based on the desired voltage reference of the load. For the system under study with no grid connection (i.e, $Vpcc=0$) such reference is found by
means of simple phasor calculations; i.e., by dividing the internal e.m.f. voltage of the MMC ($e_{v\star}=\left|e_{v\star}\right|e^{jwt}$) by the equivalent impedance ($Z'=R'+jwL'$).

\begin{equation}
x_{2\star}=\frac{\left|e_{v\star}\right|}{\sqrt{R'^2+\left(\omega L'\right)^2}}\sin\left(\omega t - \tan^{-1}\left(\frac{\omega L'}{R'}\right)\right) \label{eq:iv}
\end{equation}
where $e_{v\star}$ is the sinusoidal voltage that one wishes to apply to the load, with an amplitude that can vary from $0$ to $\frac{V_{dc}}{2}$.
The differential or circulating current reference $x_{1\star}$ is estimated by assuming low converter losses; hence, the mean value of the input $ac$ power $\left|P_{ac}\right|=\left|e_{v\star}x_{2\star}\right|$  is approximately equal to the mean value of the power at the $dc$ terminals of the MMC $\left|P_{dc}\right|=V_{dc}x_{1\star}$. Thus, $x_{1\star} $ can be expressed as 
\begin{equation}
x_{1\star} \approx \frac{\left|e_{v\star}x_{2\star}\right|}{V_{dc}}. \label{eq:idiffstar}
\end{equation}
Since $x_{1\star}$ is constant in this case, the voltage that drives it may be expressed simply by $u_{diff\star}=Rx_{1\star}$.

 Also, it is possible to calculate the fluctuations of the sum
$w_{\Sigma \star}$ and difference $w_{\Delta \star}$ of the capacitive energy stored between the upper and lower arms of the MMC (see \cite{Dyn} for more details on such equations). Thus,
\begin{align}
\Delta w_{\Sigma \star} =& \nonumber\int_{0}^t\left(-e_{v\star}x_{2\star} + \left(V_{dc}-2u_{diff\star}\right)x_{1\star}\right) dt \\
\Delta w_{\Delta \star} =& \int_{0}^t\left(\frac{x_{2\star}}{2}\left(V_{dc}-2u_{diff\star}\right)-2e_{v\star}x_{1\star}\right)dt \label{eq:fluct}
\end{align}
\begin{remark} 
A high-pass filter is needed for the term that is being integrated in equation (\ref{eq:fluct}) to leave out the power error caused by neglecting assuming no losses in the converter (\ref{eq:idiffstar}).
\end{remark}
The average value of $w_{\Sigma \star}$ and $w_{\Delta \star}$ is given as a reference from the user; typically, $W_{\Sigma \phi \star} = \frac{C}{N}V_{dc}^2$, and $W_{\Delta \phi \star}= 0$. Using the energy
estimation, it is now possible to calculate the upper and lower arm voltages.
 \begin{align}
 u_{CU\star} = \nonumber\sqrt{\frac{N}{C}\left[\left(W_{\Sigma \phi \star}+\Delta w_{\Sigma \star}\right)+\left(W_{\Delta \phi \star}+\Delta w_{\Delta \star}\right)\right]} \\
u_{CL\star} = \sqrt{\frac{N}{C}\left[\left(W_{\Sigma \phi \star}+\Delta w_{\Sigma \star}\right)-\left(W_{\Delta \phi \star}+\Delta w_{\Delta \star}\right)\right]} \label{eq:uc}
\end{align}
The remaining state variables $u_{\Sigma \star}$ and $u_{\Delta \star}$ may now be calculated as
\begin{align}
x_{3 \star} =& \nonumber u_{CU \star} + u_{CL \star}\\
x_{4 \star} =& u_{CU \star} - u_{CL \star} \label{eq:ucs}
\end{align}
In addition, the upper and lower insertion indexes are calculated by
\begin{align}
n_{u\star}=& \nonumber\frac{\frac{Vdc}{2}-e_{v\star}-u_{diff\star}}{u_{cu\star}} \\
n_{l\star}=& \frac{\frac{Vdc}{2}+e_{v\star}-u_{diff\star}}{u_{cl\star}} \label{eq:nu}
\end{align}
Finally, consistent with this methodology, the control at the desired trajectory is defined as:
\begin{align}
u_{1 \star} =& \nonumber n_{u\star}+n_{l\star} \\
u_{2 \star} =& n_{uk\star}-n_{l\star} \label{eq:mmcu}
\end{align}
\subsection{The Controller}
To summarize, the controller equations are:
\begin{align*}
\dot z&=-y\\
u&=-K_p y+ K_i z + u^\star\\
y&=\frac{1}{2}
\begin{bmatrix}
x_{1\star} x_{3}-x_{1}x_{3\star}+\frac{1}{2}x_{2\star}x_{4}-\frac{1}{2}x_2x_{4\star}\\
x_{1\star}x_{4}-x_{1}x_{4\star}+\frac{1}{2}x_{2\star}x_{3}-\frac{1}{2}x_2x_{3\star}\end{bmatrix}
\end{align*}
with reference variables $x_{2\star}$ defined in \eqref{eq:iv}, $x_{1\star}$ in \eqref{eq:idiffstar} and $x_{3\star}$, $x_{4\star}$ in \eqref{eq:fluct}, \eqref{eq:uc} and \eqref{eq:ucs}. Also, $u_{1 \star}$
and $u_{2 \star}$ are derived in \eqref{eq:mmcu} and \eqref{eq:nu}.
\subsection{Simulations and experimental results}

The MMC simulation scenario has been set up in Matlab/Simulink using a high efficiency model \cite{EfficientModel} to test the validity of the control. The considerations are the following: the converter has
$2N=10$ submodules, $N$ in each arm (upper and lower). The input DC voltage is $V_{dc}=150 V$, the reference voltage $e_{v\star}$ has an amplitude of $\frac{V_{dc}}{2}$ and the frequency set to $50Hz$. The
frequency of the balancing algorithm \cite{Ondyn,Dyn,MMC} that balances the $N$ capacitor voltages is set to $20kHz$. The internal capacitance, resistance and inductance are respectively set to $C=3.3mF$,
$R=8\Omega$ and $L=10mH$. The load resistance and inductance values, respectively, are $R=6\Omega$ and $L=20mH$.

\begin{figure}[t!]
\captionsetup[subfigure]{labelformat=empty} \centering \subfloat[Simulation Trend.]{\includegraphics[width=0.48\linewidth]{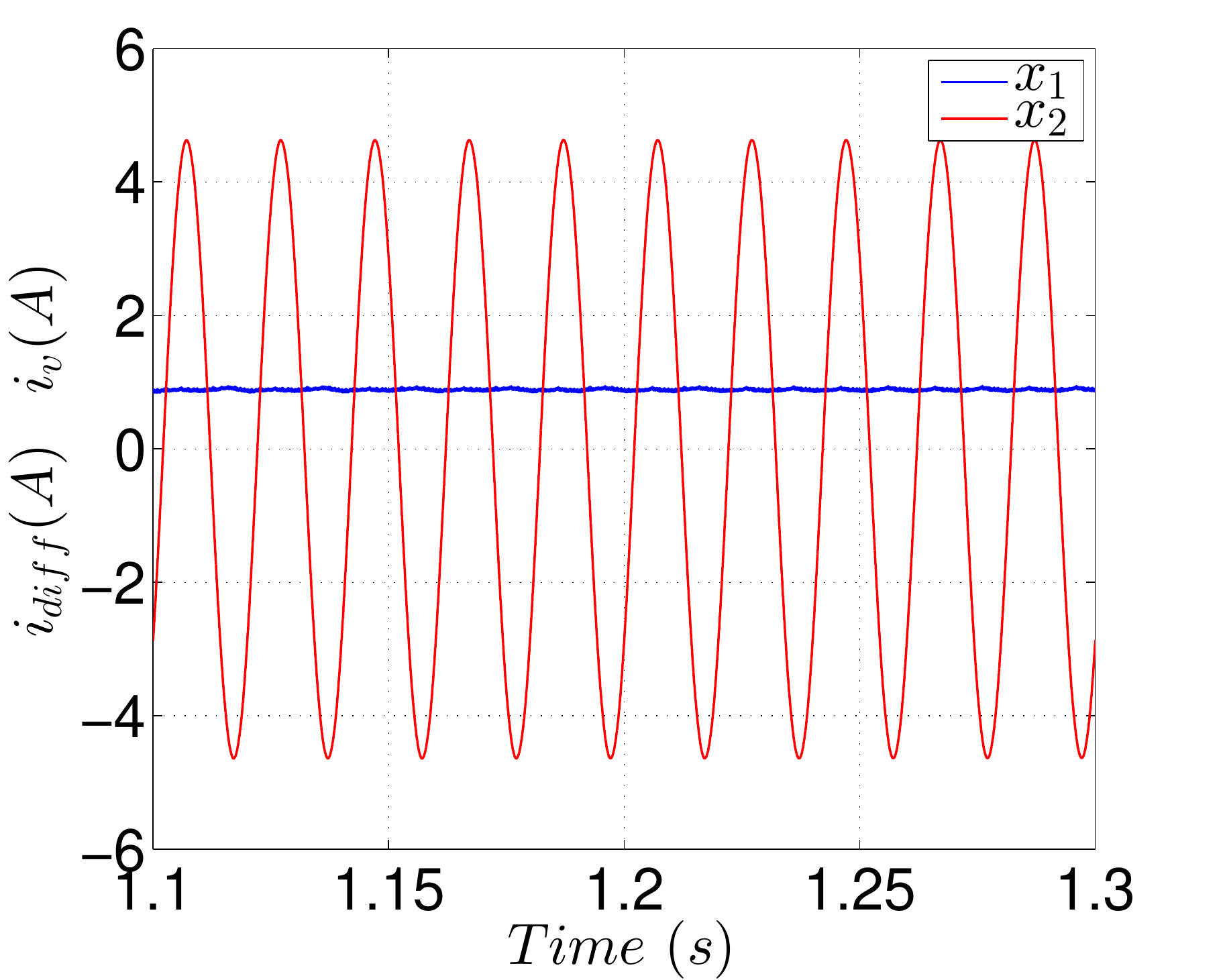}}\quad \subfloat[Experimental
Trend.]{\includegraphics[width=0.48\linewidth]{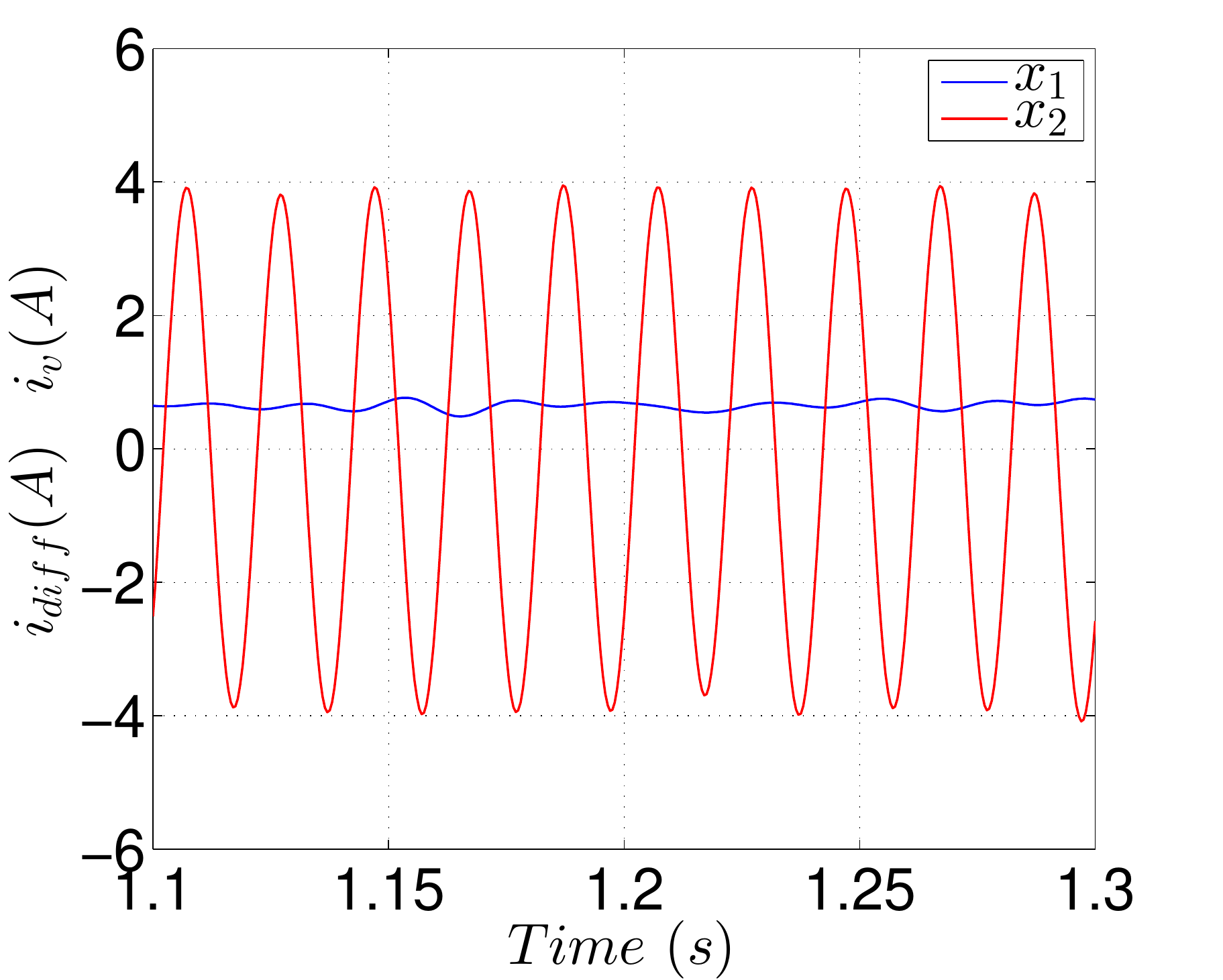}} \caption{Differential and Load currents.} \label{fig:x1x2} \subfloat[Simulation
Trend.]{\includegraphics[width=0.48\linewidth]{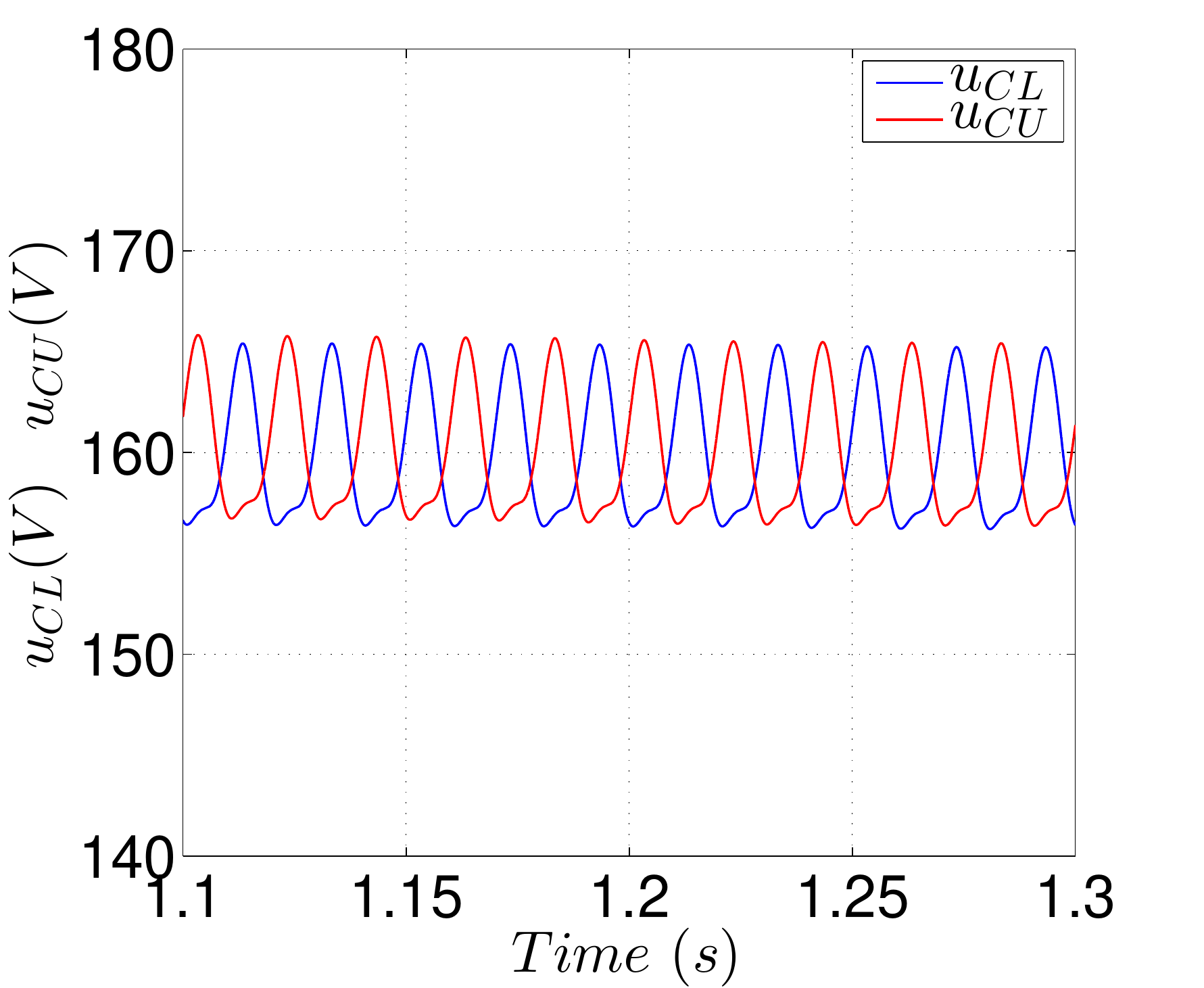}}\quad \subfloat[Experimental Trend.]{\includegraphics[width=0.48\linewidth]{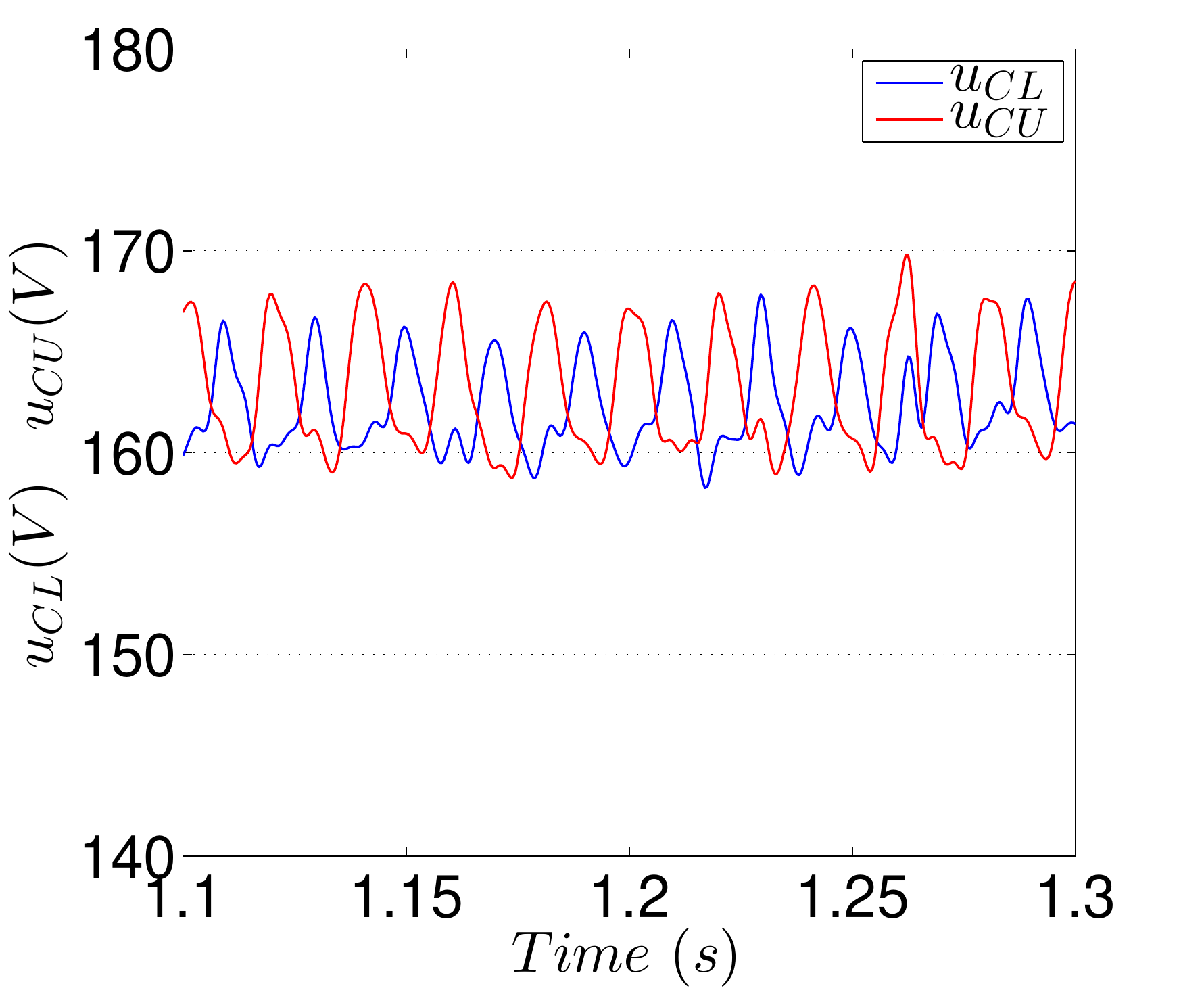}} \caption{Upper and Lower capacitor
voltages.}\label{fig:x3x4} \subfloat[Simulation Trend.]{\includegraphics[width=0.48\linewidth]{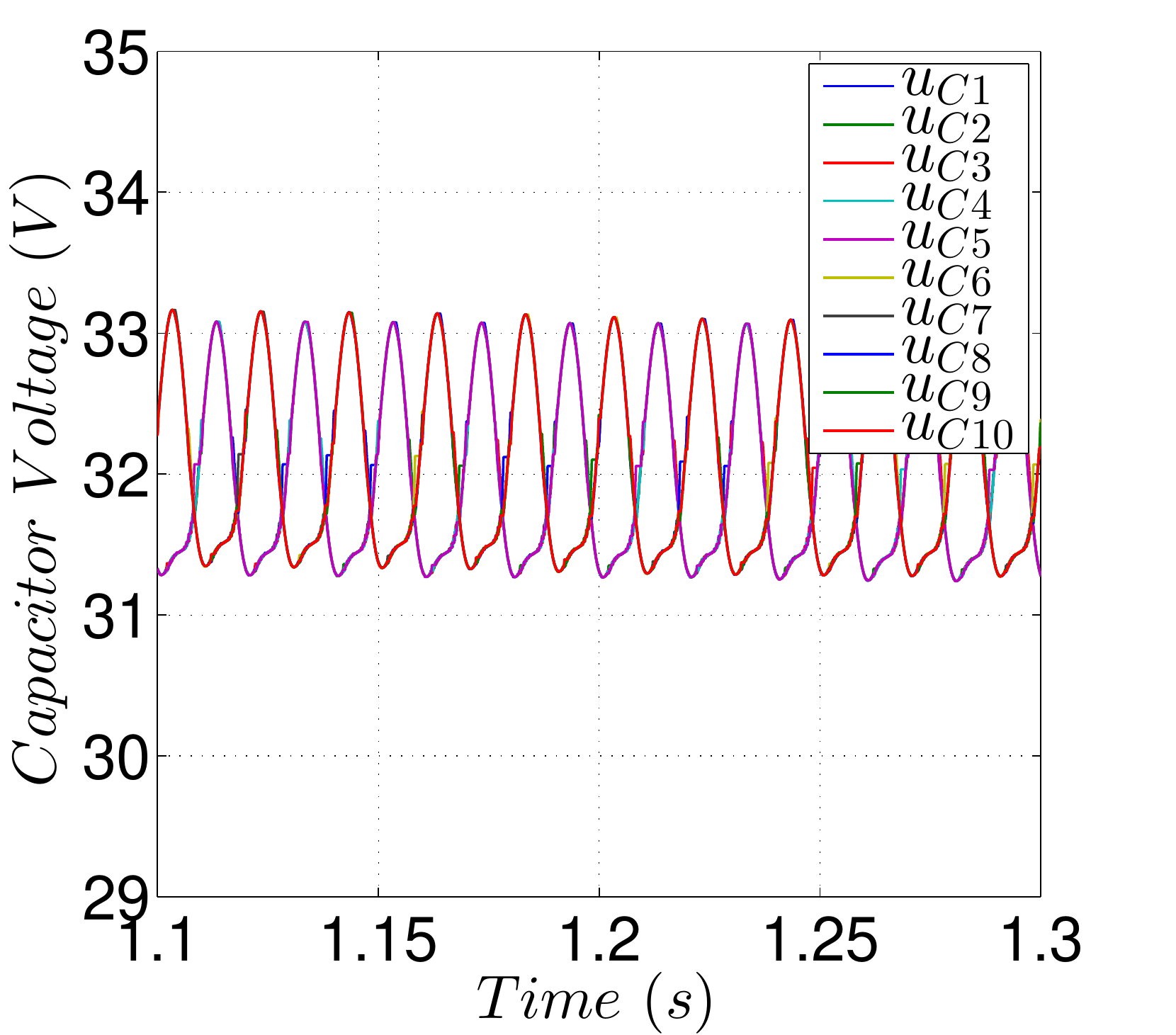}}\quad \subfloat[Experimental Trend.]{\includegraphics[width=0.48\linewidth]{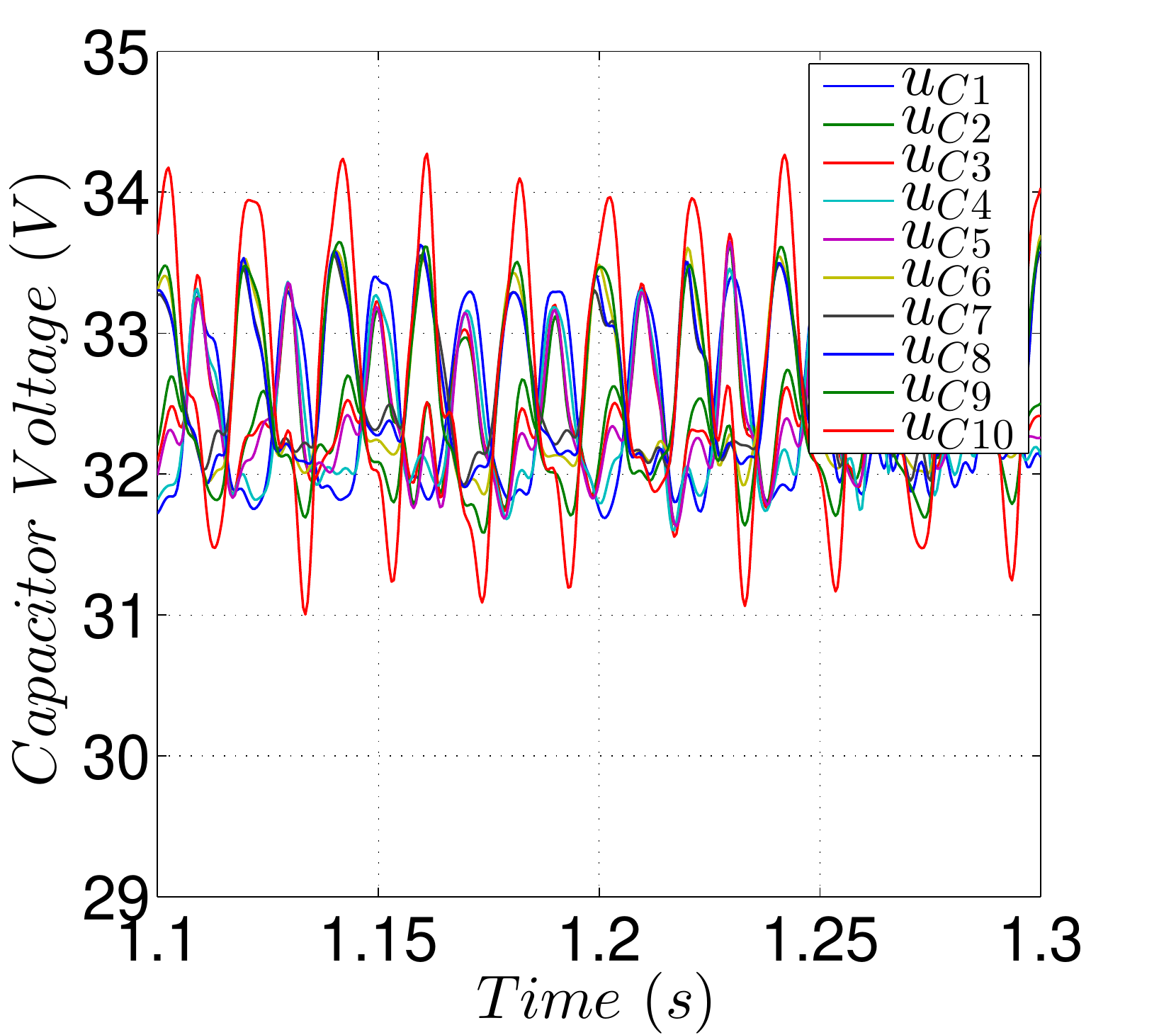}}
\caption{Individual capacitor voltages.} \label{fig:uc10} \subfloat[Simulation Trend.]{\includegraphics[width=0.48\linewidth]{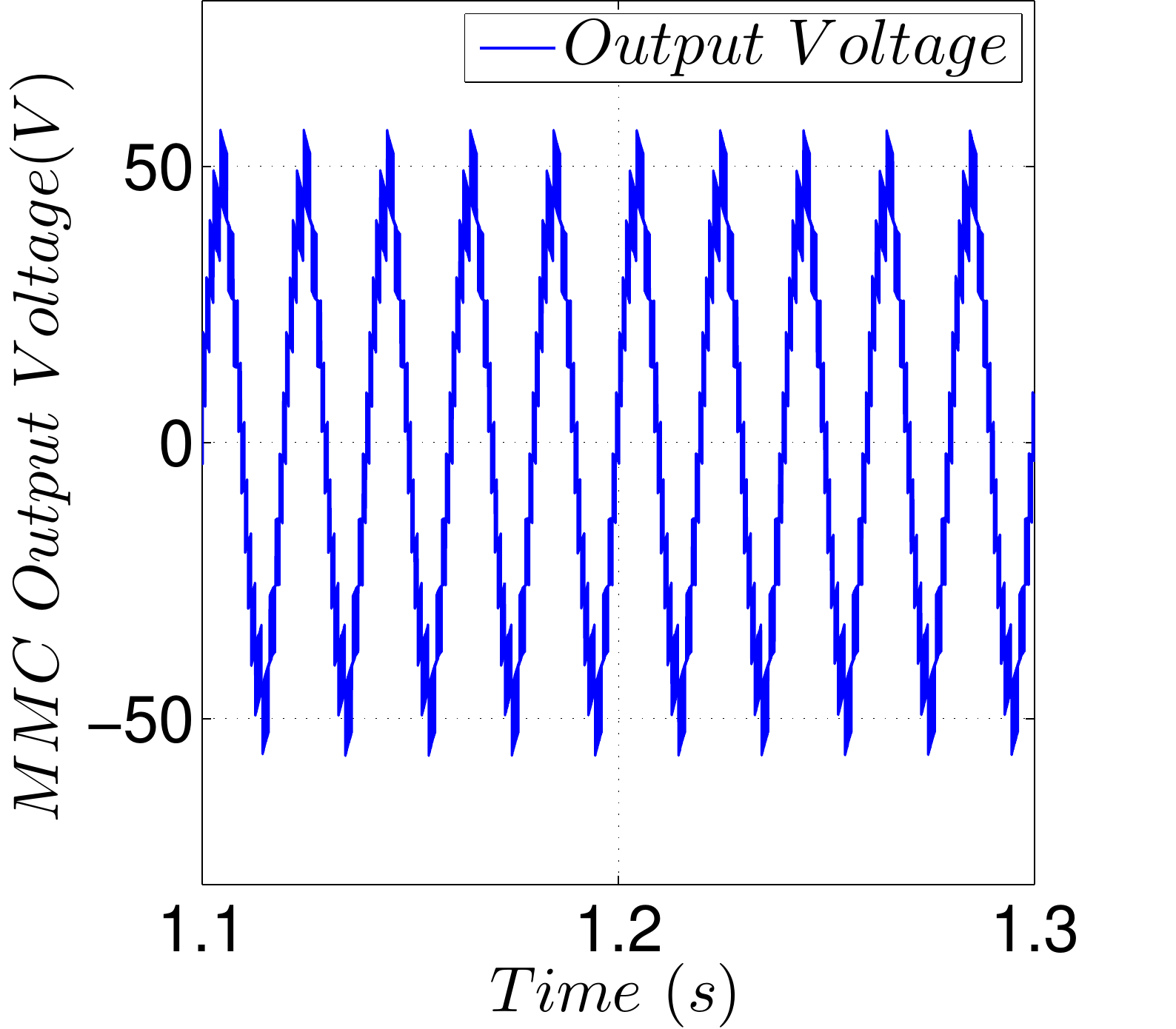}}\quad \subfloat[Experimental
Trend.]{\includegraphics[width=0.48\linewidth]{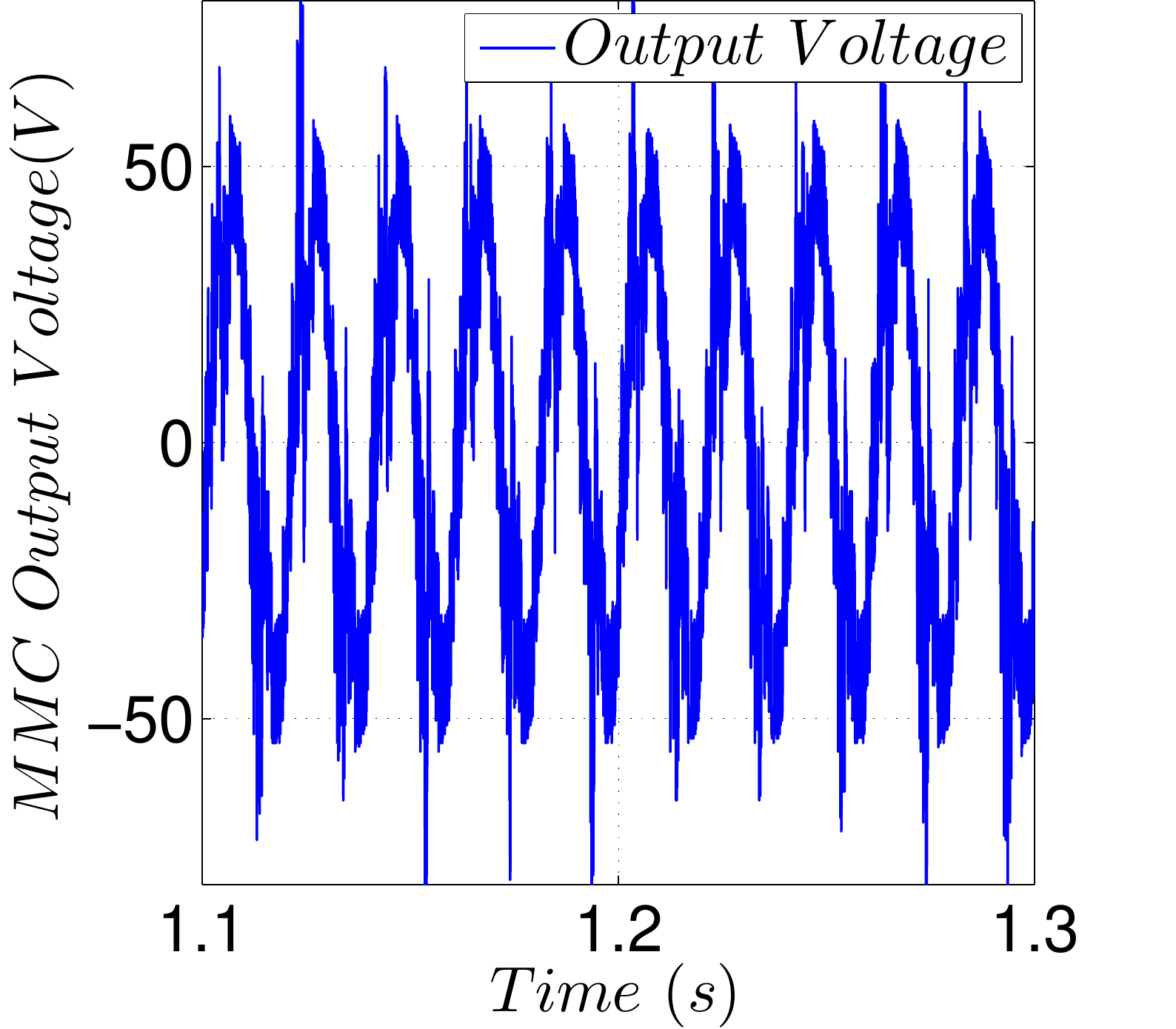}} \caption{Multilevel output waveform applied to the load.} \label{fig:vmmc}
\end{figure}

The figures presented show the controller performance for both  the simulation and implementation scenarios. Figure \ref{fig:x1x2} shows the differential and load current, respectively,  $i_{diff}$ and $i_{v}$.  
The grid current tracks with significant accuracy its reference orbit. The differential current and the capacitor voltages achieve good performance; however, they are strongly influenced by high-order harmonic pollution. This is caused by the long switching dead-time requirement of the experimental setup. In Fig. \ref{fig:x3x4} it is shown the values for the upper and lower capacitor voltages $u_{CU}$ and $u_{CL}$. According to our calculations in the simulations, there is an error of
around $3\%$ between these voltages signals and their references.  It can be concluded that is a direct consequence of the approximations made in the estimation process. Finally, in Fig. \ref{fig:uc10} are depicted the $2N$ voltages of
the MMC capacitors while fig. \ref{fig:vmmc} shows the output multi-level waveform of the converter.

\section{Conclusions}
In this article the trajectory tracking problem for power converters based on passivity foundations has been solved. Because of the use of passivity (of the incremental model) the controller is a simple PI. The
stability results are global and hold for all positive definite gains of the PI. In fact, this outcome extends previous results obtained for the regulation case.

 The performance of the controller was tested by means of some realistic simulations and experiments for both the boost and the Modular Multilevel Converters models. A future work within this area of research includes the use of observers for partial state feedback and its succeeding application to power converters.

\bibliography{cisneros_cep14}
\end{document}